\documentclass[preprint, 3p]{elsarticle}
\journal{Solar Energy Materials and Solar Cells}

\usepackage{amsmath}
\usepackage{hyperref}

\begin{document}
	\begin{frontmatter}
	
	\title{Theory of random nanoparticle layers in photovoltaic devices\\applied to self-aggregated metal samples}
	\author[CSIC]{C. David\corref{c1}}
	\ead{christin.david@csic.es}
	\author[UPVLC]{J. P. Connolly}
	\author[UPVLC]{C. Chaverri Ramos}
	\author[CSIC]{F. J. Garc\'ia de Abajo}
	\author[UPVLC]{G. S\'anchez Plaza}
	
	\cortext[c1]{corresponding author}
	\address[CSIC]{Instituto de Qu\'imica-F\'isica Rocasolano - Consejo Superior de Investigaciones Cient\'ifcas, Serrano 119, 28006 Madrid, Spain}
	\address[UPVLC]{Nanophotonics Technology Center, Universidad Polit\'ecnica de Valencia, 46022 Valencia, Spain}

\begin{abstract}
	Random Al and Ag nanoparticle distributions are studied on varying substrates, where we exploit the nanosphere self-aggregation method (NSA) for fabrication. Relying on the measured particle size distributions of these samples, we develop a theoretical model that can be applied to arbitrary random nanostructure layers as is demonstrated for several distinct NSA samples. As a proof of concept, the optical properties of the exact same particles distributions, made from the quasi random modeling input with electron beam lithography (EBL), are investigated from both theory and experiment. Our numerical procedure is based on rigorous solutions of Maxwell's equations and yields optical spectra of fully interacting randomly positioned nanoparticle arrays. These results constitute a new methodology for improving the optical performance of layers of nanoparticles with direct application to enhanced photovoltaics.
\end{abstract}

\end{frontmatter}

\section{Introduction} 
\label{sec.intro}
	Metal nanoparticles (MNPs) possess promising optical properties, which improve light-matter interaction in a number of applications \cite{Schuller2010,Polman2008,Yeshchenko2009,Zhang2008,Polman2012}. For example, these MNPs enhance light coupling in solar cells by means of efficient scattering \cite{Ferry2009,Atwater2010,Deceglie2012}, making them well suited to third generation photovoltaic devices. More generally, plasmon-assisted efficiency enhancement of solar cells may involve two classes of phenomena:
	
	(a) For collective  effects, a  resonance  condition  determined  by  the  geometry  of  the  plasmonic structure and its environment is realized. Such resonances include localized surface plasmons (LSPR) and propagating surface plasmon polaritons (SPP). The former leads to high local field enhancement the vicinities of the MNPs. Light driven processes such as multiple exciton generation \cite{Gregg2003, Nozik2008} and photoluminescence \cite{Yuan2011} can be amplified as a result of this effect. Additionally, localized SPPs and propagating plasmons, can be excited in wave\-guide structures within the solar cells \cite{Pala2009}.
	
	(b) Scattering effects are observed for all frequencies and may be optimized over the whole optical spectrum, increasing the optical path length within the solar cell. Increased scattering and a high plasmon mode density enhances the effective absorptivity by efficiently coupling the light to the underlying structure \cite{Catchpole2008}. These phenomena are particularly suitable to thin solar cells technologies\cite{Ferry2010}.

Building on these ideas, the European FP7-248909-LIMA project \cite{LIMA} develops  state-of-the-art techniques \cite{Yuan2011, Kerschaver2006} in order to improve industry compatible light management in solar cells on the nanoscale. This is achieved by developing design rules of random plasmonic particle layer (PPL) specifications which may yield an overall increase in absorption efficiency. The PPL is combined with standard antireflection coating (ARC) technology, yielding improved performance and acting similar to cell texturing, as a result of light scattering at the PPL.

The solar cell design rules developed to date have concentrated on regular structures and optimized absorption efficiency for a narrow spectral range {\cite{deAbajo2007}}. Random particle layers, however, are known to provide a broad spectrum of overlapping plasmon modes \cite{Aydin2011}. Light coupling efficiency increases with the number of modes that can be coupled into the system \cite{Chau2005}. However, theoretical studies and predictions of the optical properties of random patterns are difficult \cite{Gresillon1999,Bozhevolnyi2002}. Most research in this direction considers alloys or structured metal films \cite{Shalaev1998,Nakayama1984} describing their optical properties with effective medium theory. 

In this context, dielectric particles are as well investigated for the potential use of their photonic properties for photovoltaics \cite{Grandidier2011,Akimov2010}.

In this article, we use a simple route towards the modelling of optical spectra of random samples where only the particle size and distance distributions are needed as an input for optical calculations including interactions between these particles.
We use this formalism to simulate the optical properties of specific random distributions of nanoparticles. Note, that the theory of random distributions for photovoltaic applications can equally be applied for metal and dielectric structures. However, we concentrate on Al and Ag random layers.

In the following discussion, section \ref{sec.NSAsamples} describes the experimental industry-compatible nanosphere self-aggregation method (NSA) of PPL fabrication and characterisation. This yields experimental random distributions of particle geometries which are however difficult to model because of uncertainties in the actual spatial arrangement of the particles.
\begin{figure}[t!]\centering
	\includegraphics[width=0.8\textwidth]{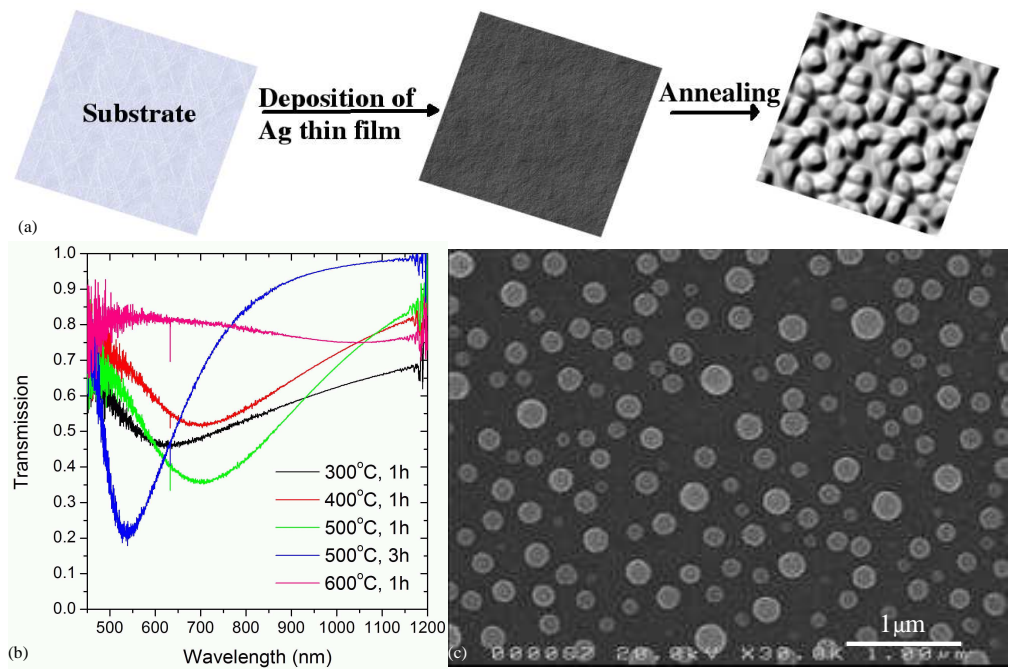}
	\caption{(a) Schematic process of the nanosphere self-aggregation (NSA) method. (b) Optical characterization results for different annealing temperatures and times. This demonstrate how sensitive the optical properties are to fabrication parameters that result in different statistical characteristics of the NSA samples. (c) SEM image of a sample fabricated at 500$^\circ$C and annealed for 3 hours for an initial Ag film thickness of 20 nm. }
	\label{fig.NSAsamples}
\end{figure}

To resolve this issue, we present in section \ref{sec.REBLsamples} a combined experimental and theoretical study on the same exact particle distributions, chosen with a high degree of randomness. First, an exact quasi-random experimental distribution is constructed based on mimicking a truly random NSA distribution. This is achieved by electron beam lithography (EBL), following specific particle sizes and positions that are analytically defined. Then, the theoretical response is modeled for this quasi-random distribution and compared with the experimental data for the same exact distribution.

The combination of the analytically exact experiment and theory is a novel approach. It first allows validation of the modelling methodology by comparison with exact data presented in section \ref{sec.REBLsamples}. This in turn allows the modelling of experimental self-aggregated particle distributions, yielding a methodology for optimizing experimental PPL properties.

NSA samples fabricated at both high and low temperatures are investigated. We discuss in detail our results in section \ref{sec.results}. Our conclusions derived from this work are given in section \ref{sec.conclusion}. An overview on abbreviations used throughout this article is provided in section \ref{sec.abbrev}.

\section{Nanosphere self-aggregation}
\label{sec.NSAsamples}
A major challenge when exploiting plasmonic structures for solar cells is the production of nanoparticle layers with a cost-effective fabrication process. The nanosphere self-aggregation (NSA) method \cite{Conolly2010,Yang2010} provides a cheap, CMOS-compatible process, enabling integration of MNP layers in photovoltaic device manufacturing.

The NSA method, illustrated in Fig. \ref{fig.NSAsamples}(a), consists of annealing a thin Ag film which self-aggregates into a sheet of randomly distributed nanoparticles. The process parameters of the Ag precursor thickness, and annealing temperature and time, allow control of MNP geometrical parameters that influence particle homogeneity and density.
To achieve a high degree of homogeneity, temperatures of hundreds of degrees Celsius are favored during the annealing process. The obtained particles are then hemi-spherical with negligible differences between minor and mayor axis. However, in our case we cannot use high temperatures in order to prevent damage of the solar cell, so that we restrict ourselves to the $100^\circ$-$400^\circ$C temperature range, hence requiring 1-3 hours of annealing. Finally, the choice of the Ag precursor thickness determines the mean particle size in the sample.

This method is well suited to fabricate large area films, and as such is suitable for mass fabrication of semiconductor solar cell. Other types of solar cells, such as synthetic dyes or organic polymer cells, are equally benefiting from the addition of plasmonic particles \cite{Kim2008}. However, the MNPs need to be placed directly inside the dye or polymer, which requires mehods such as electrodeposition or chemical deposition.
 
  The resulting NSA samples are structurally characterised by scanning electron microscopy (SEM, fig. \ref{fig.NSAsamples}(c)) and by spectrally resolved transmission and reflection measurements (FTIR). As can be seen from measurements under different anneal conditions in Fig. \ref{fig.NSAsamples}(b), the final particle layer characteristics crucially determine the optical spectra. 

While the particle geometry obtained by NSA (Fig. \ref{fig.NSAsamples}(c)) is relatively homogeneous, the plasmon resonance is strongly dependents on average particle size, and the broadening is strongly related to the width of the particle size distribution. 

The optical characterization demonstrates NSA plasmonic layer transmission (Fig. \ref{fig.NSAsamples}(b))  above $90\%$ in the long wavelengths range, but also shows a reduced transmission for short wavelengths. This reduced transmission and associated loss due to reflection, interference, and absorption, has been identified as a concern in prior art, as the losses at short wavelength range \cite{Yang2010} compete with the enhanced absorption efficiency at long wavelengths. Solutions reducing the short wavelength loss include adjusting particle size and distribution, and incorporating an additional index matching layer on top of the PPL  \cite{Munday2011}.
We numerically study the optical performance of random PPLs by applying exact electrodynamic modelling \cite{deAbajo1999} briefly summarised in the next sections. 

The resulting design methodology enables us to identify desirable particle diameters for the highest possible scattering efficiencies. The recommendations from the modeling are well within the wide range of parameters achievable with the NSA technique, and distributions with particle sizes of less than $100$ nm in diameter are best suited. In this range, MNPs are efficient scatterers and dispersive losses are negligible \cite{Catchpole2008, Yang2010}. A detailed image analysis provides us with information about the particle size and distance distribution. With the knowledge of these statistical parameters, we intend to predict optical spectra of related random patterns as obtained by the NSA method using the numerical procedure presented next.
\begin{figure}[t!]\centering
	\includegraphics[width=0.8\textwidth]{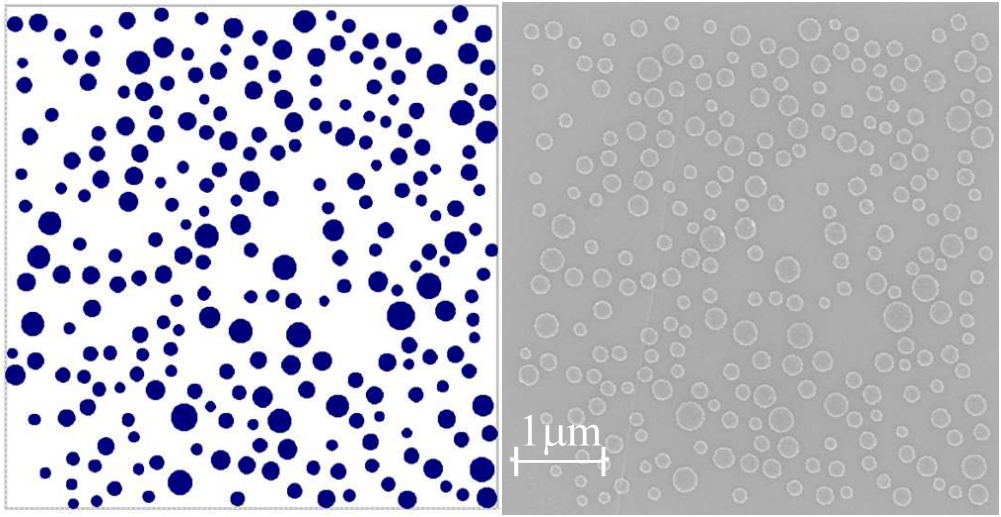}
	\caption{Comparison of a numerical sample (left) and the corresponding actual sample (right, SEM picture) fabricated with electron beam lithography (EBL), covering an area of $5\times5\mu$m$^2$ with homogeneous Al particles. The achieved agreement is excellent, thus enabling direct comparison of optical characterization and its simulation.}
	\label{fig.REBLsamples}
\end{figure}

\section{Random e-beam lithography}

\label{sec.REBLsamples}
Exact theoretical modeling of random PPLs is challenging. We introduce a quasi-random electron beam lithography (REBL) technique that enables direct comparison between experiment and modeling. EBL is an accurate nanofabrication method well suited for systematic study of precisely defined geometries, enabling exact comparison with modeling. Its resolution and the particle size in the quasi-random samples guarantee that aberrations from the ideal circular shape are negligible. In this section we therefore investigate exact quasi-random distributions theoretically and experimentally.

\begin{figure*}[t]
	\includegraphics{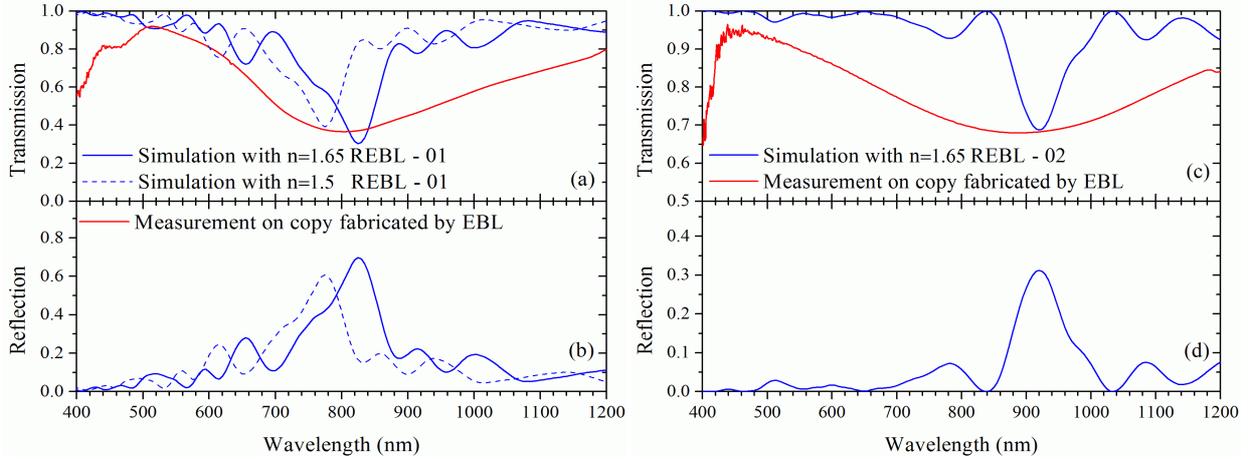}
	\caption{(a),(c) Measured (red) and calculated (blue, upper panel) transmission spectra for two $5\times5\mu\rm{m}^2$
REBL samples of Al particles. By comparing with measurements on exact copies of the numerical samples fabricated by standard EBL we show that the main plasmon peak position of the random ensemble and the transmission value thereof are well described by our computational approach. Additionally, (a) gives results for two glass environments with different refractive index n to demonstrate the effect of the permittivity of the surrounding medium. (b),(d) From the calculated reflection (blue, lower panel), eq. \eqref{eq.refl}, we obtain the transmission by neglecting absorption contributions $T=1-|r|^2$.}
	\label{fig.REBLresults}
\end{figure*}

\begin{figure*}[ht!]
	\includegraphics{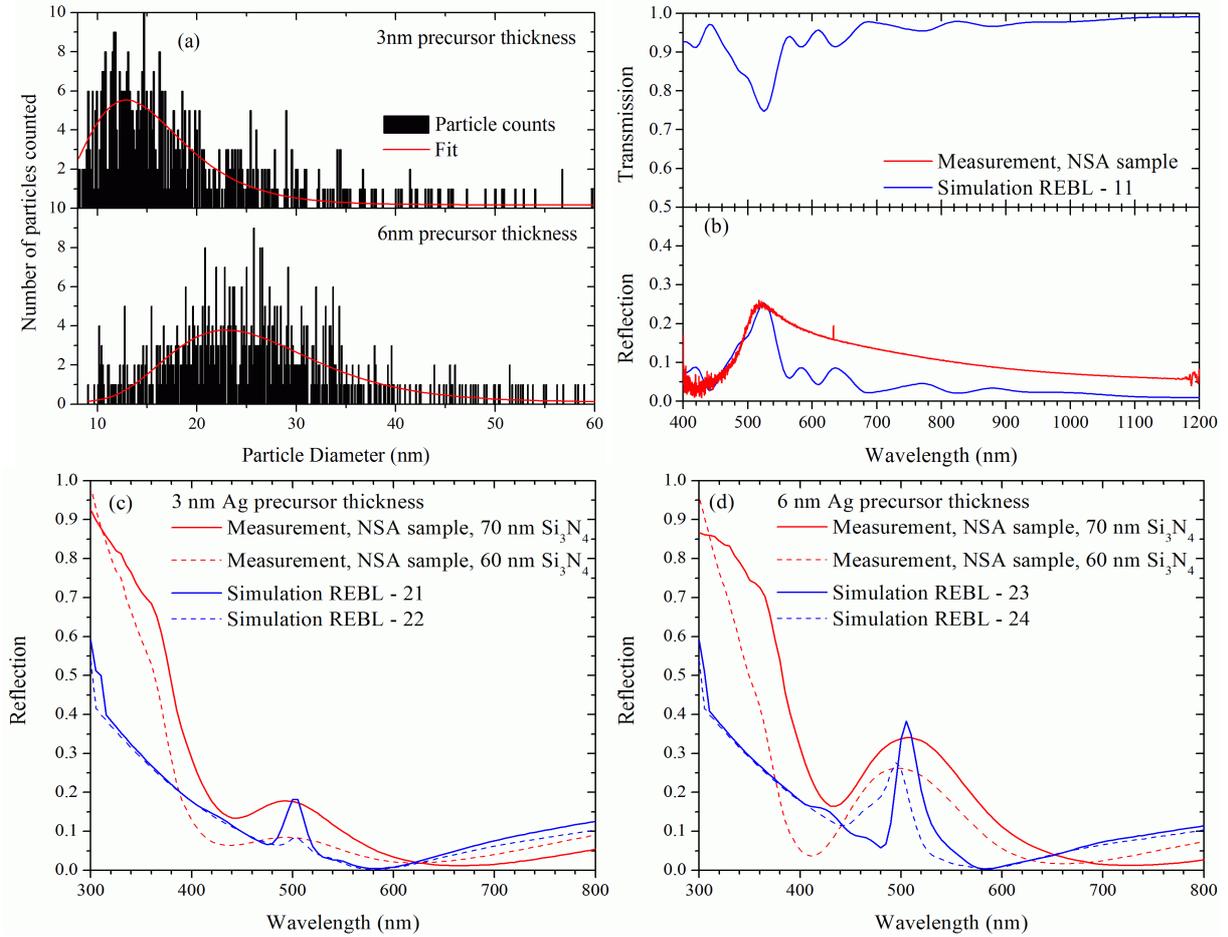}
	\caption{Examples of application of the REBL approach (blue curves show simulated spectra) to real NSA samples (measurements depicted in red) for different Ag precursor thicknesses, annealing temperature and annealing time, as well as different substrates. (a) Statistical parameters are retrieved by an automated image analysis and fitting of the particle diameter counts to a suitable distribution function (here Gaussian). (b) Reflection spectrum for a high temperature (500$^\circ$C, 3h) sample encompassing glass with a Ag precursor thickness of 20 nm. Excellent agreement of plasmon peak position and maximum reflection is found between theory and experiment. (c)-(d) Results on low temperature samples fabricated on Si$_3$N$_4$ layers of different thickness on Si. The Ag precursor thickness was varied from 3 nm to 6 nm. In the numerical samples of particle distributions, this is reflected through distinct statistical parameters, as shown in (a).}
	\label{fig.TrueNSAresults}
\end{figure*}

Numerical quasi random PPLs are obtained with a random number generator using a suitable distribution for particle sizes and, independently, a uniform distribution of particle positions. We provide several $5\times5\mu\rm{m}^2$ samples numerically, maintaining minimum distances as determined by EBL resolution. Physical Al samples with exactly the same geometries were fabricated with EBL, enabling direct comparison of optical spectra calculated and measured on equal quasi random samples.
A comparison of the exact numerical sample specification with SEMs of the EBL fabricated experimental analogue show exact agreement  (see Fig. \ref{fig.REBLsamples}). Note however that the shape of the particles fabricated through EBL is cylindrical rather thean spherical, in contrast to the assumption in the simulations. For optical measurements, a much larger area has to be provided, so we use repeated patterning of the random sample to obtain areas of $200\times200\mu\rm{m}^2$.
Parallel experimental and computational optical characterization (Fig. \ref{fig.REBLresults} and Fig. \ref{fig.TrueNSAresults} and following discussions) provides optimum insight into the physical behaviour of the truly random Ag PPL fabricated by NSA and the quality of the theoretical modeling. The modelling method is based on exact electrodynamic modeling mentioned above, using a multiple scattering technique \cite{deAbajo1999} for a large number of interacting particles. They are represented in terms of self-determined multipolar expansions up to a maximum value of the multipolar order $\sim25$ in order to obtain numerical convergence. This method includes retardation arising from both particle size and cluster size. 

Here, each spherical particle $\alpha$ randomly placed at $\vec r_\alpha$ contributes to the electric field $\vec E$ via its corresponding far-field-amplitude $\vec f_\alpha$. This contains the full optical response of the particle $\alpha$, including the refractive index of the environment and the self-consistent interaction with all other particles. The particle interaction is calculated on the basis of the scattering matrix of the single particles, and thus, it includes all spectral and geometrical information. The far field can thus be written
\begin{align}
 \vec E(\vec r) = \sum_\alpha \vec f_\alpha \frac{e^{i k|\vec r-\vec r_\alpha|}}{|\vec r-\vec r_\alpha|}.
 \label{eq1}
\end{align}
The wavevector is defined as $\vec k=(\vec Q, k_z)$, with $k_z$ perpendicular to the 2D array and $k_z=\sqrt{k^2-Q^2}$. The particle positions are $\vec r_\alpha=(\vec R_\alpha,0)$, assuming that the relative $z$-coordinate does play a minor role due to the small particle sizes. Thus, we can express eq. \eqref{eq1} as
\begin{align}
 \vec E(\vec r) &= \sum_\alpha \vec f_\alpha \int \frac{d\vec Q}{(2\pi)^2} \frac{2\pi i}{k_z}e^{-i\vec Q\vec R_\alpha}e^{i\vec k\vec r}.
 \label{eq2}
\end{align}
Both the numerical and the experimental samples consist of a $5\times5\mu\rm{m}^2$ region that is repeated periodically over a square array of $40\times40$ periods of spacing $d=5\mu\rm{m}$. The sum in eq. \eqref{eq2} can then be separated into a sum over each unit cell (index $\alpha_0$ of the contributing particles) and a sum over all unit cells (index $j$), so that the positions become $\vec R_\alpha=\vec R_{\alpha_0}+\vec R_j$. This can be performed for an infinite array by separating it as
\begin{align}
	\sum_\alpha \vec f_\alpha	e^{-i\vec Q\vec R_\alpha}	&= \sum_{\alpha_0} \vec f_{\alpha_0} e^{-i \vec Q \vec R_{\alpha_0}}\sum_{\vec R_j} e^{-i\vec Q\vec R_j}.
\end{align}
Using $\sum_{\vec R_j}e^{-i \vec Q \vec R_j}=\frac{(2\pi)^2}{A}\sum_{\vec G}\delta(\vec  Q + \vec G)$ and performing the 2D discrete Fourier transform in the particle plane for the regular superlattice, we find
\begin{align}
\vec E(\vec r) &= \sum_{\alpha_0} \vec f_{\alpha_0} \sum_{\vec G} \frac{2\pi i}{A k_z^{\vec G}} e^{i\vec G\vec R_\alpha}e^{i \vec k^{\vec G} \vec r},
\end{align}
where we now sum over reciprocal lattice vectors $\vec G$ of the superlattice and the wavevector has become $\vec k^{\vec G} = (-\vec G, k_z^{\vec G})$ and $k_z^{\vec G}=\sqrt{k^2-G^2}$. Here, $A$ denotes the area occupied by the particles in the unit cell (i.~e., $A=25\mu\rm{m}^2$). Then, the total far-field amplitude $\vec f$ for  $(z\rightarrow\infty)$ of the PPL is used to calculate the transmission and reflection coefficients. A high angular resolution in the measurements assures normal scattering conditions $\vec G=0$. We can write (for normal incidence)
\begin{align}
\vec E(\vec r) &= \vec f \sum_{\vec G} \frac{2\pi i}{Ak_z}e^{ik_z|z|},\qquad\vec f=\sum_{\alpha_0}\vec f_{\alpha_0},
\end{align}
and find the reflection coefficient at a given wavelength $\lambda=\frac{2\pi}{k_z}$ to reduce to
\begin{align}
\label{eq.refl}
	r &\approx \frac{i (\vec f\cdot\hat{\vec x})\lambda}{A}\frac{n}{E^{\rm inc}}.
\end{align}
Here, $n$ indicates possible contributions from spots of higher order $\vec G\neq0$. We assume $n=1$ for our calculations.
We note that the procedure is identical for metallic or dielectric particles.

In contrast to regular patterns of MNPs, which display a narrow plasmon resonance, a PPL exhibits a broad plasmon resonance which can be understood as a superposition of localized resonances from individual particles as well as resonances emerging from interparticle coupling. The host permittivity, in which the PPL is embedded, is included as a wavelength independent constant. In Fig. \ref{fig.REBLresults}(a), we investigate the influence of the permittivity of the surounding medium. Optical measurements of physical REBL samples are directly compared with optical simulations in Fig. \ref{fig.REBLresults} giving valuable feedback on the procedure.

Finally, after analyzing Al particle sizes and distributions in NSA samples, numerical samples using these geometrical parameters are modeled and characterized. An improved prediction of expected transmission (calculated as $T=1-|r|^2$ and thus not showing absorption explicitly) and reflection spectra is then obtained (see Fig. \ref{fig.TrueNSAresults}). Wavelength-dependent calculations of finite substrates, as used in these experiments, were performed separately and combined with the PPL response simulated in a homogeneous medium. A summary of the statistical parameters used can be found in Table \ref{tab.REBLparameters}, demonstrating that this procedure can cope with several hundreds of particles including their full interaction and retardation. Note, that a multiple scattering approach is much more feasible than any FDTD calculation on random samples of that size.

\section{Discussion}
\label{sec.results}
Optical spectra are presented for REBL samples in Fig. \ref{fig.REBLresults} and directly compared to the corresponding FTIR measurements on exact copies made with EBL. The $5\times5\mu\rm{m}^2$ samples on glass show very good agreement both in the plasmon peak position and in the expected minimum transmission. The broadening is however underestimated. The obtained oscillatory behavior indicates the different plasmon peak positions arising from (a) particles of different sizes and (b) the interaction between particles (coupled modes) that naturally lead to a smooth broadening of the main peak in the measured, more extended samples. We note here again, that the physical samples include larger ensembles of repeated quasi random patterns, whereas the simulations are restricted to a single pattern, see Fig. \ref{fig.REBLsamples}. The effect of the host permittivity is shown in Fig. \ref{fig.REBLresults}(a). As expected, for higher permittivities the main peak is shifted to higher wavelengths. Note that the drop in the measured transmission for wavelengths below 450 nm (red curves) is due to the lower FTIR detector sensitivity in that spectral range.

\begin{table}%
\tiny
\begin{tabular}[t]{l| c|c|c|c}
							&		Mean Diameter	&	Variance	& Particles	 & Substrate						\\\hline
REBL-01				&		150 nm		&		25 nm		& 	102 & SiO$_2$\\
REBL-02				&		150 nm		& 	25 nm		&		128 & SiO$_2$\\\hline
REBL-11				&		100 nm 		&		20 nm		&		250 &	SiO$_2$\\\hline
REBL-21				&		12 nm 		&		2 nm		&		555 &	Si$_3$N$_4$ \\
REBL-22				&		12 nm 		&		4 nm		&		527 &	Si$_3$N$_4$\\
REBL-23				&		24 nm 		&		10 nm		&		441 &	Si$_3$N$_4$\\
REBL-24			 	&		30 nm 		&		10 nm 	&		710 &	Si$_3$N$_4$\\\hline\hline
\end{tabular}
\caption{Parameters of particle distributions used as input for optical calculations on quasi random REBL samples. The first two samples have been fabricated with EBL using Al particles (cf. Fig. \ref{fig.REBLsamples} and Fig. \ref{fig.REBLresults}). A minimum distance compatible with the experimentally achievable spatial resolution was set to $d_{\rm min}=50$ nm. Further REBL samples assume $d_{\rm min}=1$ nm to allow for all possible spacings. NSA samples are fabricated using Ag particles. The third sample was made under high temperature NSA conditions ($500^\circ$ C), whereas the lower samples on Si$_3$N$_4$ were produced at low temperatures ($200^\circ$ C).}
\label{tab.REBLparameters}
\normalsize
\end{table}

Transmission for even shorter wavelengths is expected to decrease due to two aspects: Single particle resonances (LSPRs) are found below 400 nm for Ag and Al particles and result in typical antenna reflection. In random samples, both single particle resonances and lattice (geometric) resonances are always found \cite{Atwater2010,Yang2010}. On the other hand, very small particles, as used for the samples discussed in Fig. \ref{fig.TrueNSAresults} suffer from ohmic losses at low wavelengths. Furthermore, in the case of $\rm{Si}_3\rm{Ni}_4$, the reflection of the substrate increases strongly, see Fig. \ref{fig.TrueNSAresults} (c) and (d). Therefore, the range below 400 nm is not considered here. 

Aiming at a description of physical NSA samples, we introduce in Fig. \ref{fig.TrueNSAresults}(a) typical statistical data from NSA samples used as an input for the calculations. Image analysis of SEM pictures of the $2\times2$ cm$^2$ samples was performed on $300\times300$ nm$^2$ representative areas with total particle numbers of 500 - 1500. This experimental statistical particle data is then fed into the model, which yields a simulation of the  physical NSA samples. 

In this manner, both high and low temperature NSA samples have been fabricated and simulated. The main difference is in particle distribution characteristics, as can be seen from Table \ref{tab.REBLparameters}. The numerical method introduced relies only on the statistical paramters and is proven to describe both cases equally well, see Fig. \ref{fig.TrueNSAresults} (b).

Samples were fabricated on $\rm{Si}_3\rm{N}_4$ layers of different thicknesses on top of Si. The substrate layers optical response has been calculated seperately, and REBL calculations are obtained assuming a wavelength-independent host material. Both spectra are then combined, and although interaction between wavelength-dependent substrate and PPL is thus not included, the agreement of plasmon peak positions and expected reflection are very good. Including this interaction might lead to improved agreement, especially at low wavelengths.

\section{Conclusion}
\label{sec.conclusion}
In summary, a simple procedure is given for describing the optical properties of random metal nanoparticle distributions on substrates as possible building blocks for photovoltaic devices. A combination of experimental and theoretical approaches is presented using standard EBL fabrication to provide exact copies of numerically provided quasi random samples. Optical characterization and electromagnetic simulation on the exact same random samples are made possible by this dual approach.

Excellent agreement on the wavelength position and the value of transmission features is achieved, although experimental broadening of the spectral features is underestimated by our calculations. This indicates that the particle density is sufficient to predict peak positions and transmission values, but the number of particles taken into account in this study is too low, as the broadened features presumably arise from coupling a vast number of particles.

Our experimental work is based on the findings of Yang et al. \cite{Yang2010} and reproduces the low reflectances found for Ag particle samples. To our knowledge, previous theoretical work on random particle layers is mostly based on effective medium theories (alloys, structured metal films \cite{Shalaev1998,Nakayama1984}) and can therefore not capture the full interaction picture given in the present study.

Our procedure can be used to predict optical spectra of specific realizations of random samples. Furthermore, it can be exploited to optimize the optical properties and therefore provide statistical parameters of an optimum, random nanoparticle layer. Both metal and dielectric nanostructures can be considered. Furthermore, since the procedure is based on the scattering matrix of simple objects, it can easily be used for studying other shapes, such as hemispheres and disks on various substrates.

The method was applied to predict optical spectra of physical random samples fabricated using the nanosphere self-aggregation (NSA) and by relying only on measured geometrical parameters as an input for the calculations. 

Our study opens a new methodology for designing optimized geometries of nanophotonic structures, and in particular, samples of nanoparticles as those discussed here to improve the efficiency of optical absorption in photovoltaics.

\section{Abbreviations}
\label{sec.abbrev}
\begin{description}
	\item[MNP] metal nanoparticle
	\item[LSPR] localized surface plasmon resonance
	\item[SPP] surface plasmon polariton
	\item[PPL] plasmonic particle layer
	\item[NSA] nanosphere self-aggregation (method)
	\item[(R)EBL] (random) electron beam lithography
	\item[FTIR] Fourier transform infrared spectroscopy
\end{description}

\section{Acknowledgment}
\label{sec.acknowlegement}
The authors would like to acknowledge Amadeu Griol and Claudio Ot\'on for their support and EBL expertise. We would like to thankfully mention financial support by the EU (FP7-248909-LIMA). C. D. further acknowledges a FPU fellowship by the Spanish Ministerio de Educaci\'on.

\bibliographystyle{model1-num-names}

\end{document}